\documentclass[useAMS,usenatbib]{mn2e}

\usepackage{graphicx}

\title[The PMS age scale]{Are pre-main-sequence stars older than we thought?}
\author[T. Naylor]{Tim Naylor\\
School of Physics, University of Exeter, Stocker Road, Exeter, EX4 4QL}
\begin{document}

\date{???}

\pagerange{\pageref{firstpage}--\pageref{lastpage}} \pubyear{2002}

\maketitle

\label{firstpage}

\begin{abstract}
We fit the colour-magnitude diagrams of stars between the zero-age 
main-sequence and terminal-age main sequence in young clusters and 
associations.
The ages we derive are a factor 1.5 to 2 longer than the commonly used
ages for these regions, which are derived from the positions of 
pre-main-sequence stars in colour-magnitude diagrams.
From an examination of the uncertainties in the main-sequence and 
pre-main-sequence models, we conclude that the longer age scale is probably 
the correct one, which implies we must revise upwards the commonly used ages 
for young clusters and associations.
Such a revision would explain the discrepancy between the observational 
lifetimes of proto-planetary discs and theoretical calculations of the
time to form planets.
It would also explain the absence of clusters with ages between 5 and 30Myr.

We use the $\tau^2$ statistic to fit the main-sequence data, but find that 
we must make significant modifications if we are to fit sequences which have 
vertical segments in the colour-magnitude diagram. 
We present this modification along with improvements to methods of calculating
the goodness-of-fit statistic and parameter uncertainties. 

Software implementing the methods described in this paper is available
from http://www.astro.ex.ac.uk/people/timn/tau-squared/.

\end{abstract}


\section{Introduction}

The ability to determine the ages of pre-main-sequence (PMS) stars is 
crucial for advancing our understanding of the early phases of stellar 
evolution.
There are two key applications.
Firstly, and perhaps most obviously, we need stellar ages if one is to 
carry our experiments such as tracing the evolution of stellar angular 
momentum, or following the fraction of stars with proto-planetary discs 
as a function of time. 
Secondly, for PMS stars, the conversion from observables such as 
temperature and luminosity into mass is highly age dependent, making 
accurate ages vital for determining the mass function.
The primary method of determining the ages required for these studies
is to compare the observed properties of PMS stars with models.
The most easily accessible observables are a star's temperature and
luminosity, since they can be measured from its colours and magnitudes.
The problem is that for the same colours and magnitudes different models 
can predict ages which differ by a factor two, and even the same models 
will predict different ages depending on which colours and magnitudes are
used.
This makes meaningful comparisons between the ages quoted in the 
literature for clusters or associations at best difficult, and often 
impossible.
It was these problems which led us to devise a model-independent age 
ordering of young clusters and associations based on their colour-magnitude 
diagrams \citep{2007MNRAS.375.1220M}.
For PMS stars the primary age diagnostic is based on the fact that stars 
fade as they get older and contract towards the main sequence (MS).
We used this movement of the sequence towards progressively fainter 
magnitudes to derive an age ordering, although to do so we also had to
measure a consistent set of distances, which we derived from the more 
massive stars which have already reached the MS \citep{2008MNRAS.386..261M}.

Whilst an age ordering such as ours is useful, for example it has showed
unambiguously that different clusters take different times to reach the same
disc fraction or angular momentum distribution, for quantitative work an
absolute scale is required.
For PMS clusters and associations there are several usable age indicators, 
each of which relies on comparing stellar properties with models.
For this reason it is best to group them according to the underlying 
physics.
First is the contraction of PMS stars as they approach the MS.
As pointed out above, and discussed at length in \cite{2007MNRAS.375.1220M} 
these ``contraction'' or ``PMS'' ages are highly model dependent, and given 
the current disagreements between the models cannot yield an absolute age
scale. 
Although most stars in a young cluster or association are in the PMS
phase, the evolution of the most massive stars proceeds so fast that
they may not only have reached the MS, but evolved beyond it.
This gives us access to two more age measures.
First, having reached the MS, stars move redwards and to higher luminosities 
away from the zero-age main sequence (ZAMS), due to the increasing helium 
content of their cores.
This movement continues until the point of core hydrogen exhaustion, when 
the star has reached the reached the terminal age MS (TAMS or MS turn-off).
Finally, after the turn-off, the post-main-sequence evolution is 
driven by the burning of heavier elements which leads to 
much more rapid movement in the CMD.
This relatively high velocity in the CMD means that post-main-sequence
evolution has the potential to give precise ages.
However, for young galactic clusters the paucity of stars in this region
of the CMD means such an age can depend on just one star, and such ages are 
rightly treated with some scepticism.
Conversely, the main-sequence evolution (from the ZAMS to the turn-off)
has a larger number of stars, but the movement is often subtle, and using the 
normal technique of simply plotting isochrones over the data leads to large 
uncertainties in age, and to questions over objectivity.
However, we have been developing a method of making objective fits to 
colour-magnitude data, which should allow us to unlock the information 
in this stage of a star's evolution.
The technique, called $\tau^2$ fitting, can be viewed as an extension of 
$\chi^2$ to data points with uncertainties in two or more observables,
and to models which are distributions (not just lines) in the data space. 

The aim of this paper is to apply the $\tau^2$ fitting technique to the
main-sequence evolution of young stars, and use the resulting ages to
create a revised age scale for PMS stars.
Surprisingly, this leads to a significantly older ages than the commonly
used contraction ages, a result which we will discuss in Section \ref{discuss}.
To derive this result we first have to update our statistical techniques
originally described in \cite{2006MNRAS.373.1251N}, since, as we discuss in 
Section \ref{stats}, the technique will not work for the isochrones we
wish to fit.
We therefore lay out the changes which need to be made by following
an example through fitting (Section \ref{fit}), testing the goodness
of fit (Section \ref{goodness}) and determining the uncertainties 
in the derived parameters (Section \ref{uncer}).
Before doing so, however, we discuss the data and models we use 
(Sections \ref{data} and Sections \ref{models}).
We deal with the effects of interstellar extinction in Section \ref{extin}, 
and the details of each cluster in Section \ref{individual}.
We draw all the results together in our discussion in Section \ref{discuss}.

\section{The data}
\label{data}

To compare a set of ages derived from MS evolution with contraction 
ages we need a sample of clusters and associations which have contraction
ages, and for each of which data are available for MS fitting.
Our sample is, therefore, based on the groups we placed in age order 
using the PMS in \cite{2008MNRAS.386..261M}.
Clearly, for each of these groups we require stars in the appropriate mass
range to show significant MS evolution, but we also require extinctions and 
reliable distance measurements.
$UBV$ photometry can provide all three of these.
First the $U-B$/$B-V$ diagram provides extinctions.
Second, the upper part of $V$/$B-V$ diagram is age sensitive, tracing 
the evolution of stars from the ZAMS to the turn-off.
Finally, in the age range of interest the
lower mass stars are still close to the ZAMS, and the sequence turns
redwards, making it ideal as a distance measure.
Furthermore, the $UBV$ photo-electric system is very consistent and 
well characterised.
However, to ensure we maintain the highest level of consistency we have
restricted ourselves as far as possible to the data of Johnson and
collaborators, primarily taken in the 1950s and 1960s.
As we shall show later, the quality of these data when combined with the
transformations of \cite{1998A&A...333..231B} is impressive, giving 
$\tau^2$ values which mean the model is a good fit to the data.
Clearly we wish to avoid PMS stars contaminating our sample at faint 
magnitudes and red colours, and so for most objects we apply a cut in 
observed $B-V$ which roughly corresponds to $(B-V)_0 < 0.0$.

Most of the datasets we use have robust uncertainties derived from comparisons
of many measurements of stars.
This presents us with a problem, as the quoted uncertainties in colour are 
always smaller than those in magnitude.
Conventional error analysis yields a correlation between, say, $V$ and
$B-V$, and in previous work we have always been careful to include that
correlation when modeling the uncertainties.
The starting point for such an analysis is that $V$ and $B$ are measured
independently, and so the uncertainties in $V$ and $B-V$ are $\delta V$
and $\sqrt{\delta V^2+\delta B^2}$ respectively.
Such an analysis also leads to the conclusion that the uncertainty in $B-V$
must be larger than that in $V$, in direct contradiction to the quoted
uncertainties for most of the data presented here.
This is because it is not photon statistics which are the driver of the
uncertainties, but changes in the transparency.
In this work, we therefore model the uncertainties as uncorrelated.

\section{The models}
\label{models}

Although we will try other models later, we begin by using 
``Geneva-Bessell'' isochrones.
For the stellar interior we follow the suggestion of 
\cite{2001A&A...366..538L}, and use the 
``basic model set'' (i.e. set ``c'') of the Geneva isochrones 
\citep{1992A&AS...96..269S}.
Temporal interpolation is a much more significant issue for post-MS 
isochrones than the PMS isochrones we have fitted in the past, as there
are sharp discontinuities in the rate of change of magnitude and colour
with time, as exemplified by the MS turn-off.
We therefore use the code provided on the website to interpolate the 
isochrones to the appropriate age.
We then convert from luminosity and effective temperature to colours and
magnitudes using the tables of \cite{1998A&A...333..231B}, assuming the 
colours of Vega are zero (though $V=0.03$).
We also use Bessell et al's colour dependent extinction vectors.

For some of the most luminous stars the gravities are rather low, and 
fall just outside the range of gravities given by \cite{1998A&A...333..231B}.
In these cases we extrapolate the models by simply setting the colour to 
that for the lowest available gravity.
In these cases a linear extrapolation would be different by less than 0.001 
mags, implying that the overall error due to the extrapolation is much
smaller than the uncertainties in colour. 

For reasons explained in Section \ref{sori} we used the Tycho-2 photometry
for $\sigma$ Ori.
In this case we have used the conversion given in \cite{2000PASP..112..961B}
to convert the Geneva-Bessell isochrones into the Tycho system.
\citep[][state that the Tycho-1 and Tycho-2 systems should be identical.]
{2000A&A...357..367H}
We used the reddening vector derived in \cite{2008MNRAS.386..261M}.

\section{Statistics}
\label{stats}

In \cite{2006MNRAS.373.1251N} we introduced a solution
to the long-standing problem of how to fit photometric data to 
isochronal models in colour-magnitude diagrams (CMDs).
Whilst fitting an isochronal model (a curve) to a set of data points may 
at first appear to be a simple $\chi^2$ problem, the facts that the
data points have uncertainties in two dimensions, and that the curve is 
smeared by binarity into a two-dimensional distribution, means a more 
sophisticated technique is required.
We have now used our solution to derive ages and distances for the
young clusters NGC2547 \citep{2006MNRAS.373.1251N} and NGC2169 
\citep{2007MNRAS.376..580J}, and consistent 
distances to some of the best-studied star-forming regions 
\citep{2008MNRAS.386..261M}.
In \cite{2009MNRAS.393..538J} we derive distances to Vel OB2 and the 
association around $\gamma$ Vel, and \cite{2008MNRAS.391.1279J} use our
technique for measuring the distance to NGC7419.

\cite{2006MNRAS.373.1251N} provide a rigorous development of $\tau^2$, 
but for the purpose of understanding the improvements we have had to 
make to the method, a relatively simple intuitive interpretation gives a
better insight into the problems.
Figure \ref{ob3b_subset} shows a typical fit of a dataset 
(shown as circled error bars) to a model (the colour scale).
The model is a simulation of roughly a million stars (including binaries)
using a specific age, metallicity, mass function and distance,
which is then sampled onto a grid in colour-magnitude space.
A fit in (for example) distance can be viewed as moving the model in the 
$y$-direction 
until one obtains the strongest overlap between the model and the data.
This overlap can be quantified for a single data point by taking the function
which represents its position and uncertainties (normally a two-dimensional 
Gaussian), multiplying it on a gridpoint-by-gridpoint basis by the grid, 
and then summing the resulting values.
If the grid is $\rho(c,m)$ (where $c$ and $m$ are the colour and magnitude 
co-ordinates respectively) and the $i$th data point and its uncertainties 
$U_i(c-c_i, m-m_i)$ (where $(c_i,m_i)$ are its co-ordinates), then 
mathematically the overlap is the integral of $U_i\rho$ over the entire
space.
The product of these integrals for all the data points will therefore 
reflect the overall overlap between the data points and the model, and
so we define a statistic 
\begin{equation}
\tau^2=-2\sum_{i=1,N}{\rm ln}\int{U_i(c-c_i, m-m_i)\rho(c,m){\rm d}c\ {\rm d}m},
\label{define}
\end{equation}
whose minimum value corresponds to the best fit.

In \cite{2006MNRAS.373.1251N} we showed that this definition will, for
models which are curves in $(c,m)$ space, and only have uncertainties
in the $m$-axis, reduce to that for $\chi^2$.
However, this is only the case if one chooses to multiply $\rho$ by a
normalisation factor which is dependent on the gradient of the isochrone.
Unfortunately this normalisation factor becomes infinite if the isochrone
is vertical, and double-valued at any magnitude at which the isochrone is 
double valued.
This means our $\chi^2$-like normalisation will fail for the CMD fitting
required here, because as one moves up the sequence towards bright magnitudes
the isochrones become vertical, 
before finally switching to a negative gradient.
Furthermore, if we wish to fit in $U-B$/$B-V$ space, the isochrones
are double-valued for certain values of $B-V$.

In what follows, we therefore develop an alternative normalisation, which
allows us to fit the data.
In doing so we expose the limitations of an approximation we made when 
calculating the probability that the data are a good fit to the model.
 
\section{Fitting the data}
\label{fit}

\subsection{The model CMD}

We must first create a probability density function to fit to the data.
As in \cite{2006MNRAS.373.1251N}, we create this by simulating stars 
over the appropriate range of masses.
For each star 
we choose a mass randomly from the Salpeter IMF, and if the star is a binary,
we assign it a companion of a mass drawn from a uniform distribution
between zero and the mass of the primary.
The stellar model then provides a luminosity, gravity and effective 
temperature for each star, which we then convert into colour and magnitude
using the appropriate bolometric corrections.
If a binary companion is so low mass, or so cool, that it does not appear in 
the models it is assigned a flux of zero.
(Note that this assignment is a change from \cite{2006MNRAS.373.1251N}, but 
has been used in all our subsequent work.)
The value of each pixel in the image is then simply the number of stars
whose colours and magnitudes lie within the pixel.
We typically simulate $10^6$ stars, and for this work have used pixels of
size 0.0025 magnitudes in each axis.
This is half the value we have used in previous work, but is necessitated by
the small uncertainties of the current data.
We find the residual effects of the placement of pixel boundaries are
much smaller ($\sim$0.005 mags in derived distance modulus) than the 
uncertainties in derived parameters. 

\subsection{The normalisation of $\rho$}
\label{norm}

Before proceeding further we must address the normalisation of the
model image, $\rho$.
In \cite{2006MNRAS.373.1251N} we used our $\chi^2$-like normalisation
which was a function of magnitude.
Here, we instead explore the results of a much simpler normalisation, 
setting the integral of $\rho$ over the entire image to one.
This raises the question of how faint magnitudes we must integrate down to.
In fact, the strictly correct way to proceed would be to first multiply the 
image by the photometric completeness function, such that below a certain 
magnitude $\rho$ was zero, and then set the integral of what remains to one.
Such a normalisation has an interesting, though subtle implication.
When fitting for distance modulus as the distance modulus increases, there
is a decrease in the non-zero area of $\rho$, the region between the faintest 
observable absolute magnitude (as defined by the completeness function) and 
the brightest model star.
Given that the integral over the model remains one, this means the value of 
any non-zero pixel will increase, implying that $\tau^2$ will decrease, 
and hence the fit improve.
This is actually the correct behaviour since it means that a model which 
fully populates the upper part of the sequence is better than one which does 
not.
Practically, for our data, we can use a simpler normalisation, where
we make the integral between the faintest and the brightest data points one.
This means we have thrown away one possible source of information, but in
practice this does not significantly affect the fits.

Comparing the results obtained using this normalisation with that used in 
\cite{2006MNRAS.373.1251N} 
simply changes the values of $\tau^2$ in a given $\tau^2$ parameter
grid by an additive factor; it does not change the best-fitting parameters.
This is at first a surprising result since we are changing the value of 
$\rho$ in one part of the isochrone compared with another, which may 
appear as a weighting of the points.
However, it should be remembered that adding the logarithms of the integrals 
in Equation \ref{define} is equivalent to multiplying them together, 
so changing the relative values of $\rho$ as a function of magnitude is a 
normalisation, not a weighting process.
The only possibility for altering the best fit is if the length scale for 
changes in $\rho$ is small compared to the size of an error bar. 
Then data points will drag the fit so that they lie in the higher-valued regions
of $\rho$.
Since the data point is more likely to originate in the higher-density part
of the model this would again be the correct behaviour.

Finally, it is important to note that we have no longer ``normalised out'' the
mass function as we did in \cite{2006MNRAS.373.1251N}.
Changing the mass function will change the value of $\tau^2$.
In practice we have chosen to fix it such that $\rm{d}N/\rm{d}M \propto -2.35$,
which results in good fits to the models.

\subsection{The normalisation of $U$}
\label{norm-cmd-U}

At the same time as considering the normalisation of the model, we should 
also consider that of the uncertainty function ($U$ in Equation \ref{define}).
In $\chi^2$ fitting this is set such that the maximum value of $U$ is always
the same, so the highest probability attainable is always the same, 
corresponding to a perfect fit, i.e. $\chi^2=0$.
This is the normalisation we adopted in  \cite{2006MNRAS.373.1251N}.
However, there is another obvious possibility, setting the integral of $U$ to
be one.
This would have a very significant advantage in cases where the error
bars seem to have been significantly under-estimated, and 
to obtain a good fit (i.e. a value of $\tau^2$ which 
corresponds to a $\Pr(\tau^2)$ of approximately 0.5) one has to add an
extra uncertainty to $U$, in addition to those from the observations.
This could well be due to mis-matches between photometric systems.
In such cases the procedure we have previously adopted has been to 
calculate $\tau^2$, and then
$\Pr(\tau^2)$ for increasing values of the added uncertainty, until 
$\Pr(\tau^2$) exceeds 0.5.
However, if one normalises $U$ such that its integral is one, then conceptually
one is comparing a model which includes the uncertainties with data points 
which are $\delta$-function.
One can, therefore, simply adjust the values of the uncertainties until 
one obtains the lowest value of $\tau^2$.

This normalisation has an additional conceptual advantage.
In the case where the uncertainties are very small one can now approximate
$U$ as a 2-dimensional $\delta$-functions.
This effectively removes the integral in Equation \ref{define}, and means
one can evaluate $\tau^2$ by simply multiplying together the values of $\rho$
at the positions of the data points.

We will refer to a normalisation where the integral of the models and the
integrals of the uncertainty functions are all one as the natural
normalisation.
This clearly distinguishes it from the $\chi^2$-like normalisation used in
 \cite{2006MNRAS.373.1251N}.

\subsection{The fit}
\label{the_fit}

\begin{figure}
  \begin{center}
    \includegraphics[width=2.5in]{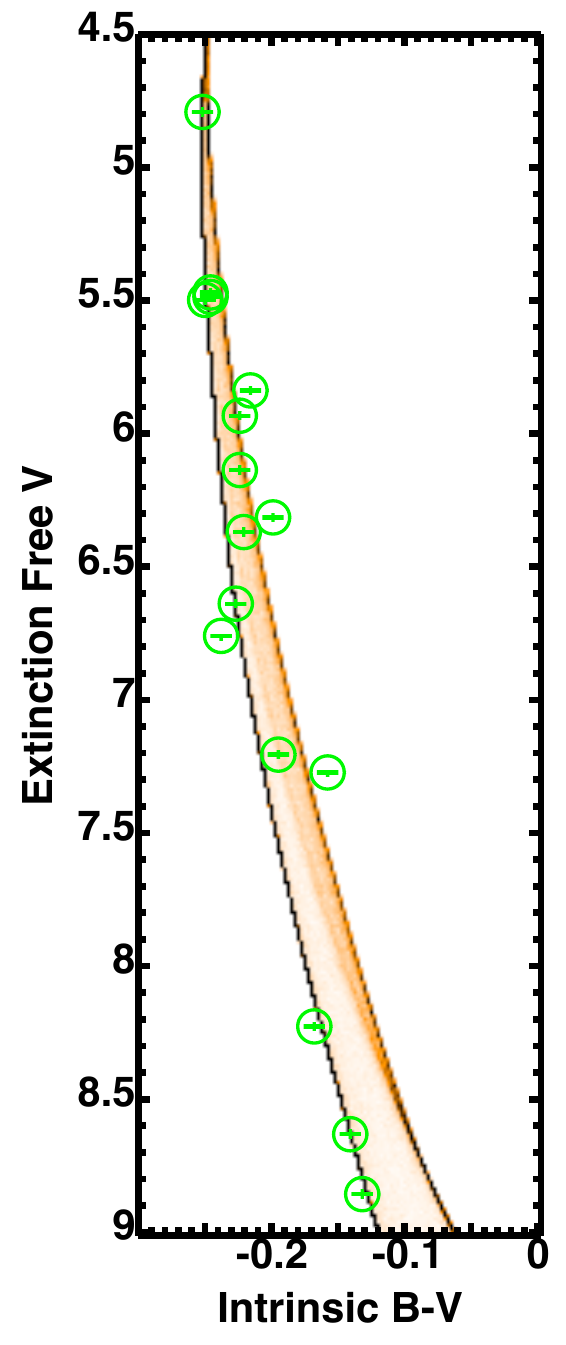}
  \end{center}

  \caption{The data and best-fitting model for 
Cep OB3b.
The colour scale is the model ($\rho$ in Equation \ref{define}) and 
the encircled error bars are the data. 
}
  \label{ob3b_subset}
\end{figure}

\begin{figure}
  \begin{center}
    \includegraphics[width=3.0in]{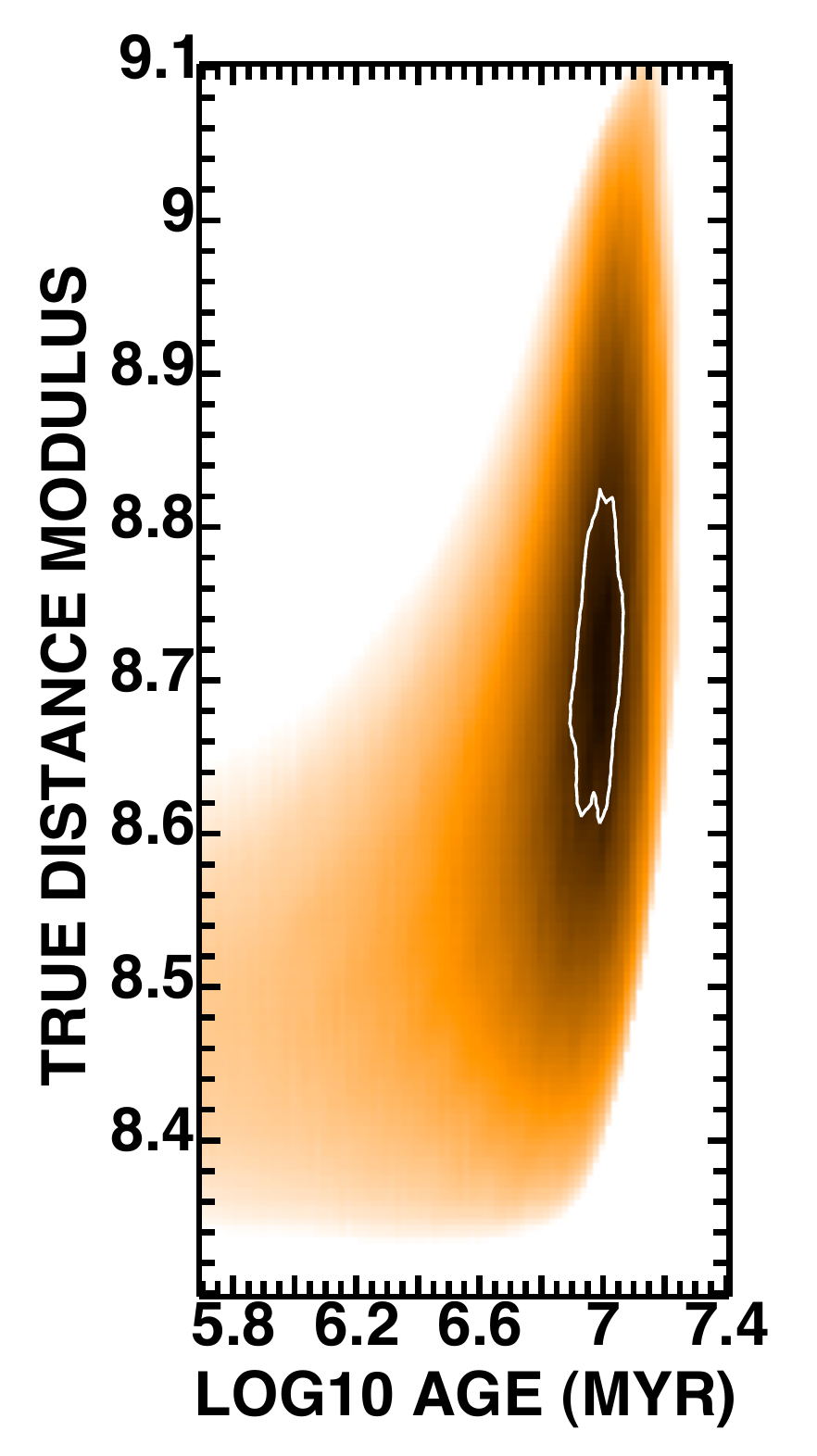}
  \end{center}

  \caption{The $\tau^2$ grid for 
Cep OB3b.
The contour is at the 68 percent confidence level.}
  \label{ob3b_grid}
\end{figure}

Given we now have the correct normalisations we can now fit our example 
data, which is a sample from 
Cep OB3b described in detail in Section \ref{ob3b}.
We calculated the extinction on a star-by-star basis
as described in Section \ref{extin}, and
after correcting for it, searched in both age and distance modulus,
evaluating Equation \ref{define} at values of our
fitting parameters which cover the range of interest.
The resulting $\tau^2$ space is shown in Figure \ref{ob3b_grid}. 
The best fit, which lies at 10 Myr and a true distance modulus of 8.7 mags,
is shown overlayed on the data in Figure \ref{ob3b_subset}.

In some fits we find that there are data points which clearly do not lie
on the sequence, and are presumably non-members.  
To deal with these objects we first fit the data with a variant of the
``soft clipping'' first described in Section 7.1 of \cite{2006MNRAS.373.1251N}.
We adapt this to the new normalisation by imposing
a maximum $\tau^2$ for any one data point.
The value used is the minimum value of
$\tau^2$ amongst all the data points, plus a fixed value, normally 20.
We implement this by calculating the probability corresponding to the 
imposed maximum $\tau^2$ and adding this to the calculated probabilities 
for each data point before calculating their $\tau^2$ values.
We then performed a second fit removing the data points which had $\tau^2$ 
values close to the clipping limit, with no clipping limit applied.

\section{Is the model a good fit?}
\label{goodness}

To test whether the model is a good fit, one must evaluate the chance of
obtaining a given $\tau^2$ or below.
One does this by calculating $\Pr(\tau^2)$, the cumulative distribution of the 
expected value of $\tau^2$.
In \cite{2006MNRAS.373.1251N} we showed how to calculate this, for no free 
parameters, in such 
a way that it was insensitive to an incorrect choice of mass function.
We then suggested that one allow for free parameters by
multiplying the $\tau^2$ axis of the distribution by $(N-n)/N$.

\begin{figure}
\begin{center}
\includegraphics[width=3.3in]{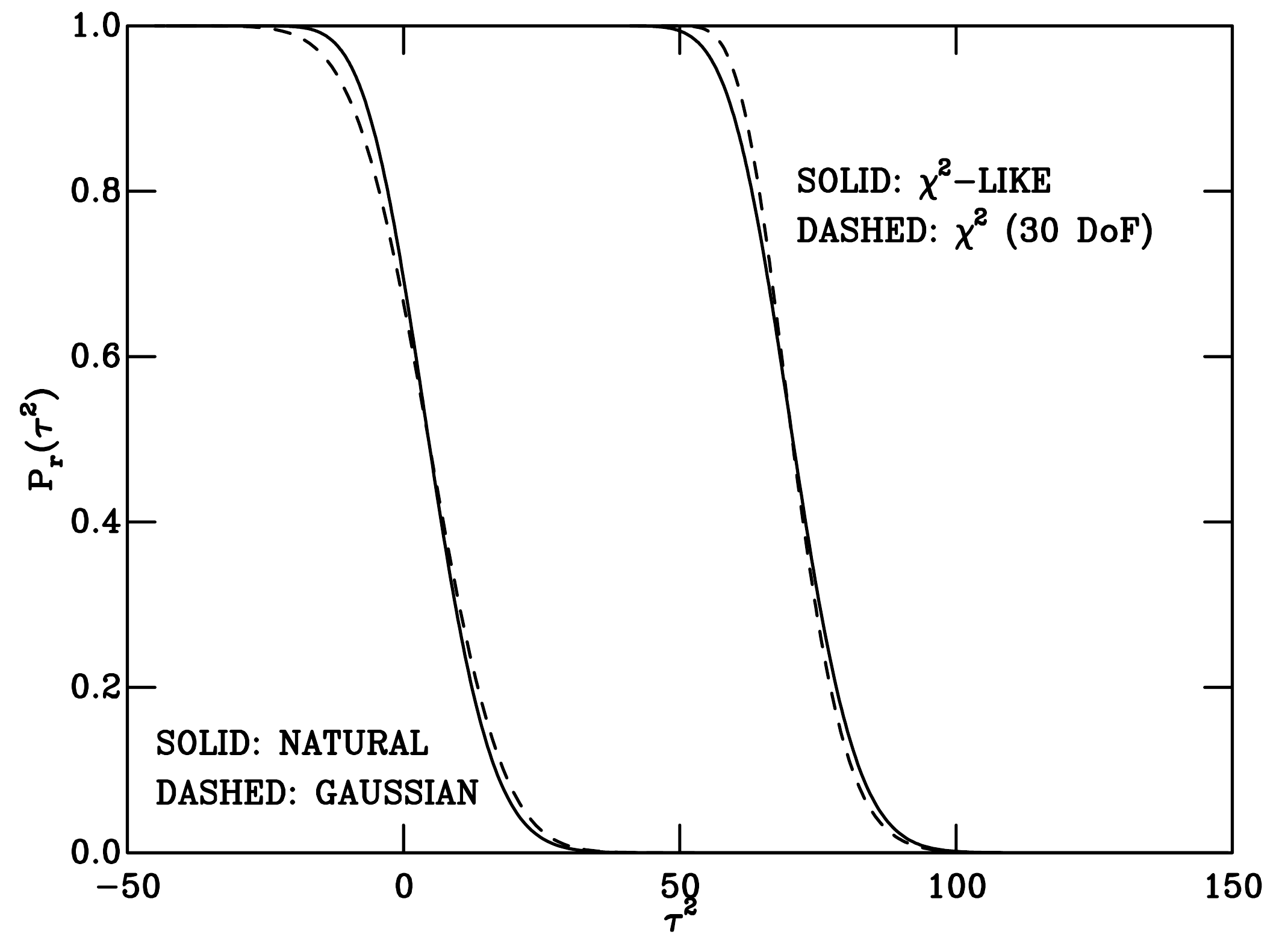}
\end{center}
\caption{
The probability of obtaining a given $\tau^2$ for a fit of 30 data points
to a main-sequence.
This is the distribution one would obtain if one created a large number
of datasets at a given distance modulus and extinction, and then ``fitted''
the data with the distance modulus and extinction fixed at their original
values.
The right-hand solid curve is for the $\chi^2$-like normalisation, 
the left-hand solid curve for a natural normalisation where the sum of the 
probability over all colours is independent of magnitude.
For comparison the dashed curves show the $\chi^2$ distribution for 30 
degrees of freedom and the
Gaussian distribution for $\sigma=60$, with their expectation
values shifted to match those for the $\chi^2$-like and natural distributions
respectively.
}
\label{tau_diff}
\end{figure}

Although our numerical simulations in  \cite{2006MNRAS.373.1251N} 
showed that the
above approach to the free parameter problem may be 
approximately correct for the $\chi^2$-like normalisation, it is 
straightforward to show that it cannot be correct in an arbitrary 
normalisation, such as the one described in Section \ref{norm}.
Consider a plot of the cumulative distribution of $\Pr(\tau^2)$
as a function of $\tau^2$ (Figure \ref{tau_diff}).
Changing the normalisation of the model means multiplying $\rho$ in
Equation \ref{define} by a constant.
This has the effect of adding a constant to the 
values of $\tau^2$ as shown by the solid curves in Figure \ref{tau_diff}.
Allowing for free parameters by scaling the $\tau^2$ axis of the distribution
by $(N-n)/N$ would yield different values for the decrease in $\tau^2$ when
adding extra parameters, depending on the normalisation.
This cannot be correct, the decrease must be additive for the shape of the
distribution to be invariant for a change in normalisation.

There is an approximate solution to this problem, though, based on the
fact that for CMD fitting, the distribution of $\Pr{\rm(}\tau^2{\rm)}$ 
is similar to $\Pr{\rm(}\chi^2{\rm)}$, save a additive factor.
The reason for this is that both distributions derive from the 
distribution of probability, and therefore $\tau^2$, in the CMD plane.
For a $\chi^2$ problem this distribution is a line, smeared by a 
one-dimensional Gaussian.
For the $\tau^2$ CMD problem the distribution approximates to two sequences 
(those of single-stars and of
equal-mass binaries) smeared by a two-dimensional Gaussian.
It is, therefore, unsurprising that the resulting distributions of 
$\Pr{\rm(}\tau^2{\rm)}$ and $\Pr{\rm(}\chi^2{\rm)}$ are similar.
So we could approximate $\Pr{\rm(}\tau^2{\rm)}$ by simply using
the $\chi^2$ distribution directly.
However, for large values of $N-n$ the differential form of the $\chi^2$
distribution tends to a normal distribution whose mean is $N-n$ and whose
$\sigma$ is $2(N-n)$.\footnote{It is interesting that the differential form of
the $\tau^2$ distribution (like the $\chi^2$ distribution) tends towards 
a Gaussian.
This is  a natural consequence of the central limit theorem, since 
multiplying functions together and then taking the logarithm is 
equivalent to averaging their logarithms.
Whilst the mean and width of the distribution are problem 
dependent, this may still provide a key to the solution in more general 
cases.}
This means we can allow for $n$ 
free parameters by subtracting the expectation value from the distribution 
of $\tau^2$, multiplying the $\tau^2$ axis by $(N-n)/N)$, and adding back 
the expectation value less $n$.
We have implemented the latter approach, as it retains any asymmetry in the
distribution.
Applying this to the Cep Ob3b data results in 
a value of $\Pr(\tau^2)$ of 0.05.
This is on the margins of acceptability, but no one datapoint is
clearly discrepant.

\section{Uncertainties}
\label{uncer}

We have found a faster method for calculating the uncertainties than that
presented in  \cite{2006MNRAS.373.1251N}.
The aim of the calculation is to place a contour in the $\tau^2$
grid of Figure \ref{ob3b_grid} which represents a region within which the 
parameters lie with a given confidence.
We can derive the uncertainties by first converting the values of
$\tau^2$ in the grid into probability, and then integrating over the 
entire grid.
We then divide this into the integral of the probabilities below progressively 
higher values of $\tau^2$ to obtain the cumulative $\tau^2$ distribution.
We can then pick off values of $\tau^2$ at given confidence limits, and draw
contours on the $\tau^2$ space.

There are four practical issues which have to be solved when using this 
method.
The first is that to carry out the integral one must multiply each pixel 
by its area.
If the axes are linear then the the area of the pixels is the same, and the 
sum of the pixels will suffice, as we normalise by the integral over the whole 
area.
However, if the age axis is logarithmic the simplest method is to
multiply the probability by the age for that pixel, before performing the sum.

The second problem is the underlying assumption that the model is correct.
This means that the fitting to create the grid must be carried out
using only those data points which are consistent with the model.
So practically this means a second fit must be carried out excluding 
any points which the first fit clipped, without any further clipping (see 
Section \ref{the_fit}).
Even so, this means one fit as opposed to fitting typically 100 Monte-Carlo
datasets for the previous technique, giving a speed improvement of a 
factor of 100.

The third issue is that one must sum the grid out to infinity.
This is less demanding than it might at first appear.
For example, if fitting a single data point in one dimension with
Gaussian uncertainties, one only has to move $\pm3\sigma$ from the 
best fit to include 99.7 percent of the total probability, which 
is accurate enough for calculating a 95 percent confidence interval.
Note, however, that the  probability enclosed for a given $\sigma$
declines as the power of the number of observables measured for each
data point, and so if generalising $\tau^2$
to many dimensions one would have to act with caution.

The final issue is that the machine
precision may be exhausted for some data points towards the edge of the 
$\tau^2$ grid, where the corresponding values of $\rho$ are very close to 
zero.
For example, if the smallest representable number greater than zero
is $1\times10^{-36}$, the
highest $\tau^2$ which can be obtained is approximately 168.
Once the $\tau^2$ for any data point lies below this probability, the computer 
will calculate $\tau^2$ for the whole data set to be infinite.
To flag such points in the grid we set them to a high value of $\tau^2$
(the number of data points times the $\tau^2$ resulting from a probability
of twice the smallest representable number).
Since such a $\tau^2$ is guaranteed to return a probability of zero, this
works transparently in code which implements the new method of calculating
confidence limits.
However, we also note this number in the header of the grid file, so 
its meaning is clear if plots are made from the file. 
Applying this technique to the 
Cep OB3b
data results in the contour
shown in Figure \ref{ob3b_grid}.

For the work here, the distance is a nuisance parameter, and we need to be
able to quote an uncertainty in age alone.
We therefore integrate the probability in Figure \ref{ob3b_grid} over all 
distance modulii
at each value of the age to create a run of probability with age.
We then define a confidence limit as that region in age which
integrates to give 68 percent of the probability, and which excludes equal 
integrals of probability above and below it. 
For Cep OB3b  this gives an age range of 8.6 -- 10.9 Myr.

\section{The extinction}
\label{extin}

Now having shown how we can fit for age, we must return to the 
question of the extinction.
We follow an improved version of the two-strand approach developed in 
\cite{2008MNRAS.386..261M}.
We first attempt to fit the the $U-B$/$B-V$ data with just the reddening
as a free parameter.
We can now \citep[in contrast to][]{2008MNRAS.386..261M}, test whether the
model is a good description of the data.
If it is, we assume the extinction is uniform, and apply the derived
extinction to all the data points.
If $\Pr(\tau^2)$ is too high, we conclude the extinction is non-uniform,
and resort to deriving individual extinctions for each star by moving them
along the (colour dependent) reddening vector until they reach the single-star
$U-B$/$B-V$ isochrone.
This is essentially a modern version of the $Q$ method of 
\cite{1953ApJ...117..313J}.
As explained in \cite{2008MNRAS.386..261M} the disadvantage of this method is
that it cannot allow for the fact the star may be a binary.
This has the effect of narrowing the dereddened sequence in $V$/$B-V$ space,
hence our preference for the $\tau^2$ method where the extinction can be
shown to be uniform.

\section{Main-Sequence Ages}
\label{individual}

We can now apply our technique to the rest of our sample of clusters and
associations to derive MS ages.
Each dataset we fit is given as an (electronic only) table as summarised
in Table \ref{ages_table}, though we show the data for $\lambda$ Ori as
an example in Table \ref{example}. 

\subsection{NGC6530}

\begin{figure}
  \centerline{
    \mbox{\includegraphics[height=4.2in]{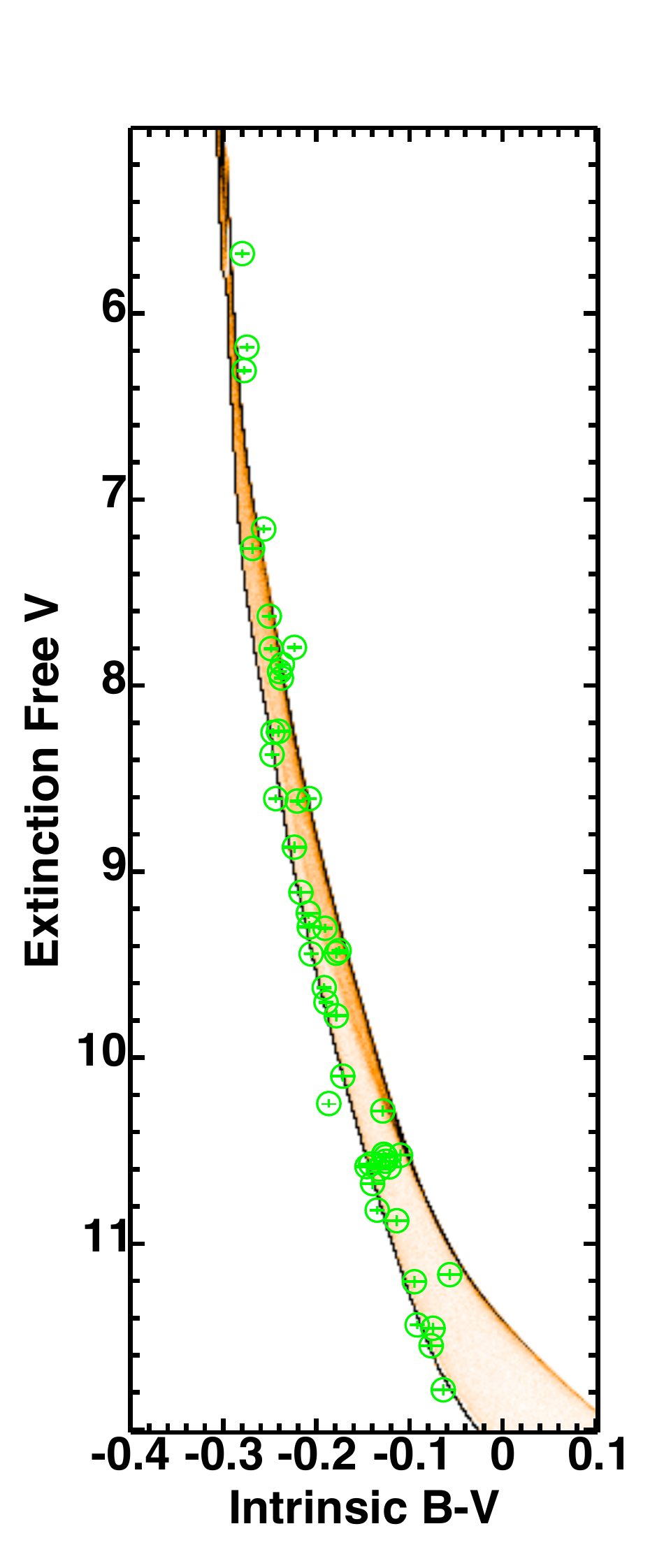}}
    \mbox{\includegraphics[height=4.2in]{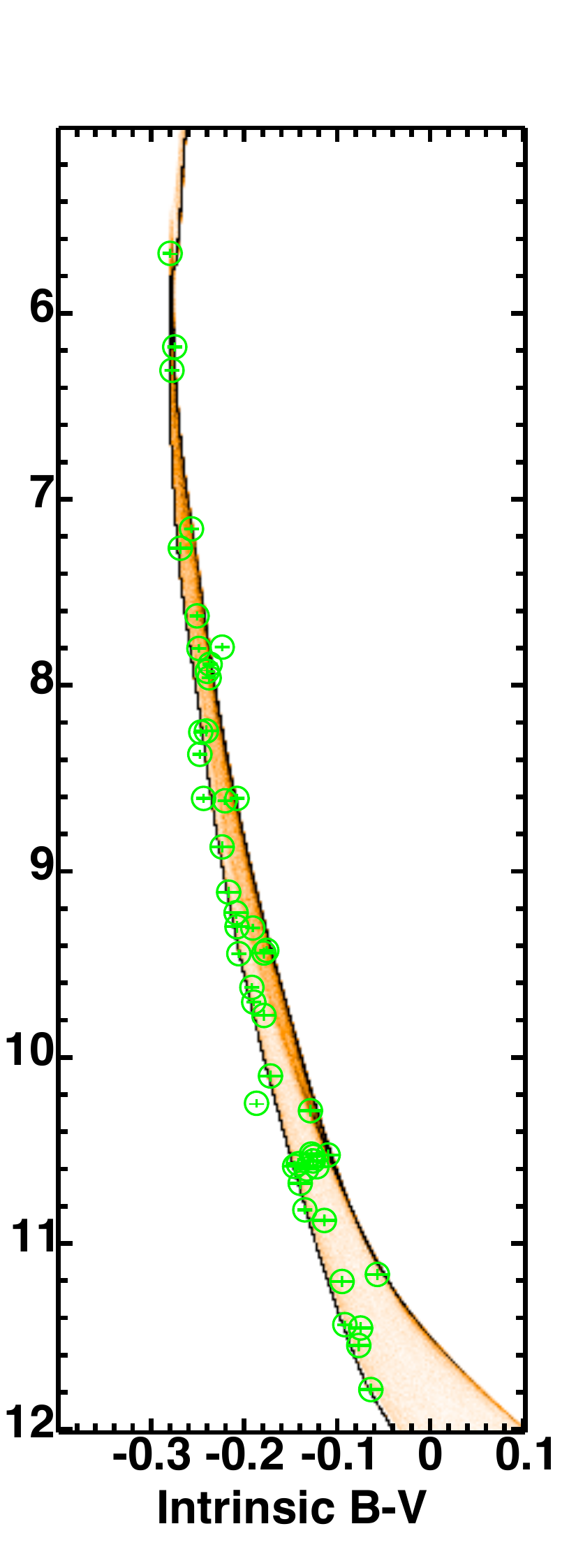}}
  }
  \caption{
The best-fitting model for NGC6530 with the age fixed at 
2Myr (left) and the best fitting model with both age and distance 
as free parameters (right).
Note how the five brightest datapoints are better fitted in the right-hand
plot, and how the group of datapoints below them would have the be
interpreted as binaries in the model on the left.
The very faintest single stars also lie marginally closer to the 
single-star sequence in the better-fitting model.
}
  \label{ngc6530}
  \end{figure}

We used the data and uncertainties of \cite{1957ApJ...125..636W}, which
are for a sample which is unbiased in colour and taken from an specific
area of the cluster.
To ensure we excluded the PMS, we selected only those stars blueward of 
$B-V$=0.28 and brighter than $V=13$ and derived individual extinctions.
The resulting fit is shown on the right-hand side of Figure \ref{ngc6530}.

\subsection{$\sigma$ Ori}
\label{sori}

Our sample consisted of the members listed by \cite{2008AJ....135.1616S} that 
are blueward of $B-V=0.1$.
We omitted the two stars noted by \cite{2008AJ....135.1616S} as variable,
and HD 37333 which is above the MS, and probably a PMS star.
We could not find a consistent Johnson $UBV$ dataset for this
cluster, and so used the Tycho-2 catalogue and its first supplement
\citep{2000A&A...355L..27H}, although it does not contain a magnitude for
$\sigma$ Ori C.
In this dataset a combined magnitude is given for $\sigma$ Ori A and B.
As $\sigma$ Ori A is itself a binary, we removed the effect of $\sigma$ 
Ori B on the combined magnitude by assuming the magnitude 
difference between the components is the mean of the values for the
difference found from 
speckle \citep{2001AJ....121.1583H} and adaptive optics 
\citep{2000AJ....119.2403T} work.
$\sigma$ Ori B will have little effect on the combined colour.

The disadvantage of using the Tycho-2 data is that we cannot determine
the extinction, as there are no $U$-band data.
We therefore simply adopted $E(B-V)=0.06$ from \cite{1994A&A...289..101B}.
The resulting $\tau^2$ contour does not close at low ages, and so we have only 
an upper limit on the age.
We therefore quote (in Table \ref{ages_table}) the upper limit below which 
68 percent of the probability lies, but exclude this cluster from further 
analysis. 

\subsection{NGC2264}

We used the photo-electric data of \cite{1956ApJS....2..365W}, 
as presented in his Table 1.
Fitting all stars blueward of $B-V=0$ for extinction in $U-B$/$B-V$ space 
gives $\Pr(\tau^2)$=0.37, implying uniform extinction over the field.
We then fitted in $V$/$B-V$ and obtained a $\Pr(\tau^2)$ of 0.06.
This is on the margins of acceptability, and there is a case that the two
data points furthest redward from the sequence should be removed.
However, in not doing so we simply enlarge our uncertainty estimate, and
so are being conservative.

\subsection{$\lambda$ Ori (Collinder 69)}

\begin{figure}
  \centerline{
    \mbox{\includegraphics[height=6.0in]{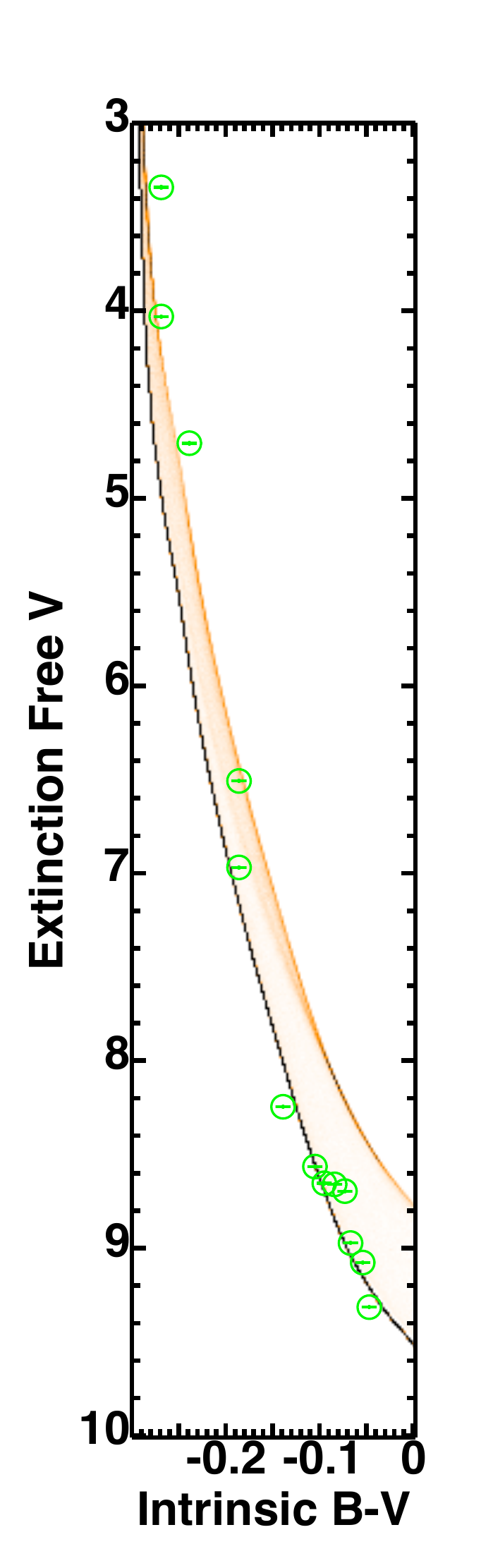}}
    \mbox{\includegraphics[height=6.0in]{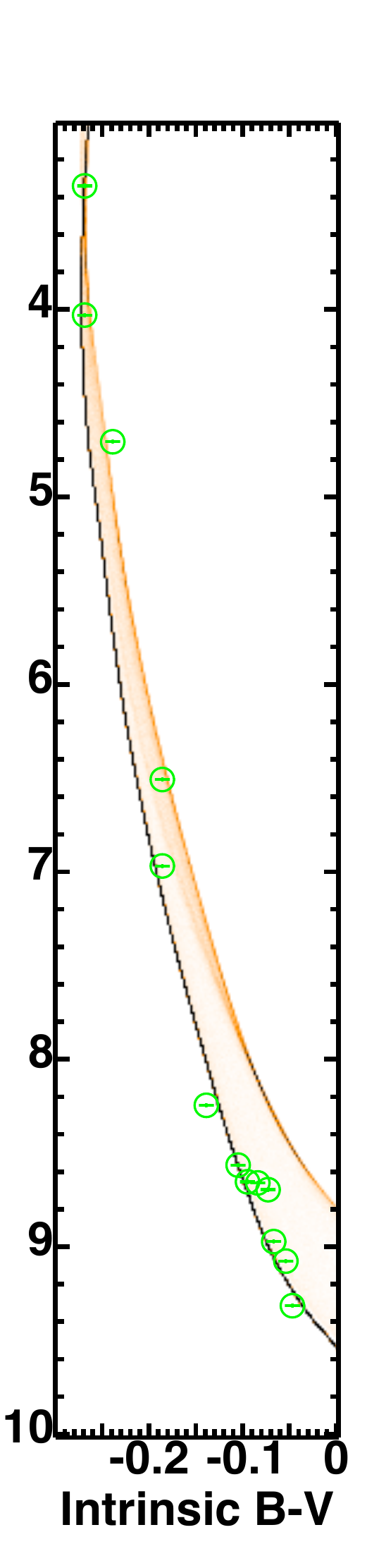}}
  }
  \caption{
The best-fitting model for $\lambda$ Ori with the age fixed at 
3Myr (left) and the best fitting model with both age and distance 
as free parameters (right).
The brightest three stars are clearly better fitted by the model on
the right.
}
  \label{lambda_ori}
  \end{figure}

We used the data from \cite{1977MNRAS.181..657M} taking only those stars 
within half a degree of $\lambda$ Ori.
We excluded objects with $B-V>0.2$, which results in a sample which is 
almost complete blueward of this colour, and has no stars redward of 
$(B-V)_0=-0.04$. 
After applying reddenings determined on a star-by-star basis, we obtained
a value of $\Pr(\tau^2)$ of 0.52, provided we assumed the uncertainties were
0.01 mags in $V$ and 0.008 mags in $B-V$ 
\citep[][do not provide error bars]{1977MNRAS.181..657M} and removed two 
objects (HD36881 and HD36913) which appear
to be non-members based on their position in the $V_0$/$(B-V)_0$ diagram.
The resulting fit is shown on the right-hand side of Figure \ref{lambda_ori}.

\subsection{NGC2362}

We used the data and associated uncertainties for NGC2362 from 
\cite{1953ApJ...117..313J}, which were taken as part of a programme to define
what became the $UBV$ system.
We used only those stars blueward of $B-V=0.04$ and excluded stars
noted as non-members by \cite{1953ApJ...117..313J}.
We also excluded the brightest star ($\tau$ CMa) as it is clearly 
beyond the turnoff.
Finally we found that star 36 gave a high $\tau^2$ in both $U-B$/$B-V$ 
and $V$/$B-V$, and star 50 in $V$/$B-V$, and so removed them from the 
fit as well.
We then measured a global extinction from the $U-B$/$B-V$ diagram, before
fitting in $V$/$B-V$. 

\subsection{Cep OB3b}
\label{ob3b}

\cite{1959ApJ...130...69B} carried out a photometric survey of stars of
spectral type A0 and earlier identified from objective prism plates.
We take the membership list from \cite{2001PhDT........75P}, but exclude 
BHJ11 for which the measurement is a combined light measurement for a 
rather wide $\Delta m = 2.5$ binary.
We de-reddened this sample on a star-by-star basis.
Using the uncertainties quoted in the paper, we obtain a just about acceptable
value of $\Pr(\tau^2)$=0.05.

\subsection{The Environs of the Orion Nebula Cluster}

\begin{figure}
  \centerline{
    \mbox{\includegraphics[height=5.0in]{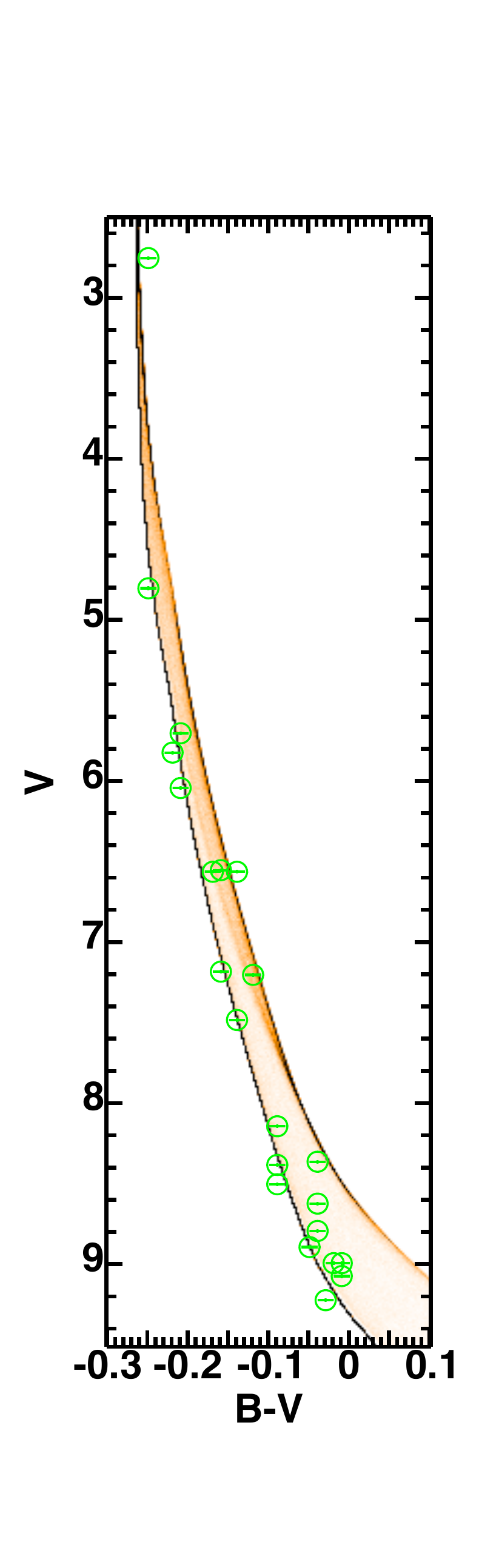}}
    \mbox{\includegraphics[height=5.0in]{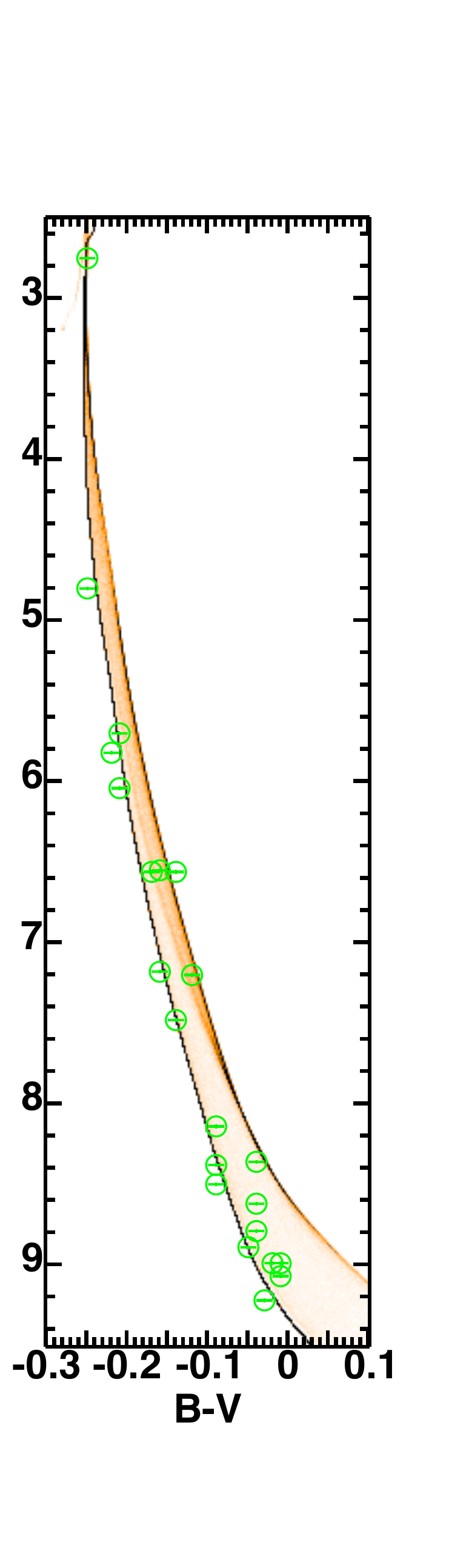}}
  }
  \caption{
The best-fitting model for the ONC with the age fixed at 
PMS age of 3Myr (left) and the best fitting model with both age and distance 
as free parameters (right).
The 68 percent confidence interval for the best-fitting age just encompasses 
3Myr, and so we expect the improvement in the fit from left to right to be
only marginal.
We can see the improvement in the fit is due entirely to the improvement 
in the fit for the brightest star, and even that is at the expense of a worse 
fit for the second brightest star.
Thus the conclusion that the statistics drive us to, that the improvement is
marginal, seems reasonable.
}
  \label{onc}
  \end{figure}

Our data and uncertainties are taken from \cite{1969ApJ...155..447W}, who
aimed to obtain photometry for as many stars as possible within the outline
of the dark cloud, since this area will be the least contaminated by
background stars.
We removed stars redward of $B-V$=0.0, those marked as variables or visual
doubles, and three stars which lie away from
the sequence in the $U-B$/$B-V$ diagram.
We fitted the data for a single extinction in $U-B$/$B-V$ space, and after
removing two outliers in $\tau^2$ obtained a good fit with $\Pr(\tau^2)$=0.46.
At first this may sound counter-intuitive, since it implies uniform
extinction, yet it is well known that the extinction of ONC members is highly
variable.
In fact it seems this only applies to stars in the central cluster.
We then fitted in $V$/$B-V$ to obtain the results in Table \ref{ages_table}.
The resulting fit is shown on the right-hand side of Figure \ref{onc}.

\subsection{NGC2547}

We took the photometry from \cite{1982AandAS...47..323C}, and 
used an uncertainty of 0.02 mags in both magnitude and colours.
This is an estimate for $U-B$ and corresponds to the deviations for
single observations derived by \cite{1982AandAS...47..323C} from comparison
with the data of \cite{1959MNSSA..18...57F}.
Although \cite{1982AandAS...47..323C} often has four observations per star we
prefer to take the view that the uncertainty represents the difference between
the photometric systems.
We use the membership list of \cite{1982AandAS...47..323C}, which is based on 
proper motions, photometry and spectroscopy, and select only stars with 
$B-V < 0.1$ to ensure we exclude PMS stars.
We found that if we included star 40, which appears to sit just below
the MS we obtained an unacceptably low value of $\Pr(\tau^2)$.
Furthermore this star is right on the edge of the proper motion distribution
of the bulk of the members, so we excluded it.

\subsection{IC2602}

We used the data of \cite{1972ApJ...173...63E}, excluding stars with $B-V>0.0$,
and HD93163, which lies away from the sequence.
Unfortunately \cite{1972ApJ...173...63E} does not provide uncertainties, but
we found a single extinction would yield $\Pr(\tau^2)$=0.79 for uncertainties
of 0.014 and 0.014 mags in $B-V$ and $U-B$ respectively, which suggests the
extinction is uniform.
Using that extinction, and an uncertainty in $V$ of 0.025 mags gives
$\Pr(\tau^2)$=0.27 when fitted in $V$ vs $B-V$.

\subsection{The Pleiades}

We again used the data and memberships from \cite{1953ApJ...117..313J}.
The $U-B$/$B-V$ diagram, especially the region where
the gradient is reversed, shows that there is variable extinction to this
cluster.
We therefore dereddened the data on a 
star-by-star basis, which limits us to $B-V<0.0$.
Before fitting we also excluded Hertzsprung 371 (which appears to be reddened).

\section{Pre-Main-Sequence Ages}

For the PMS ages we require a set of consistent ages, and we therefore
adopt the ages of \cite{2008MNRAS.386..261M}, with the following exceptions.

\subsection{NGC2547, IC2602 and Cep OB3b}

We take the PMS age for NGC2547 from \cite{2006MNRAS.373.1251N} as
38.5Myr, which is derived from isochrone fitting, though also agrees 
with the Lithium depletion age. 
The age for IC2602 (25Myr) is taken from \cite{1997ApJ...479..776S}, which 
again is based on isochrone fitting to the PMS stars.
We use a PMS age of 4.5Myr for Cep OB3b, which is from Littlefair at al (in
prep), but is based on the system of \cite{2008MNRAS.386..261M}.

\subsection{The Environs of the Orion Nebula Cluster}

\begin{figure}
  \begin{center}
    \includegraphics[width=3.3in]{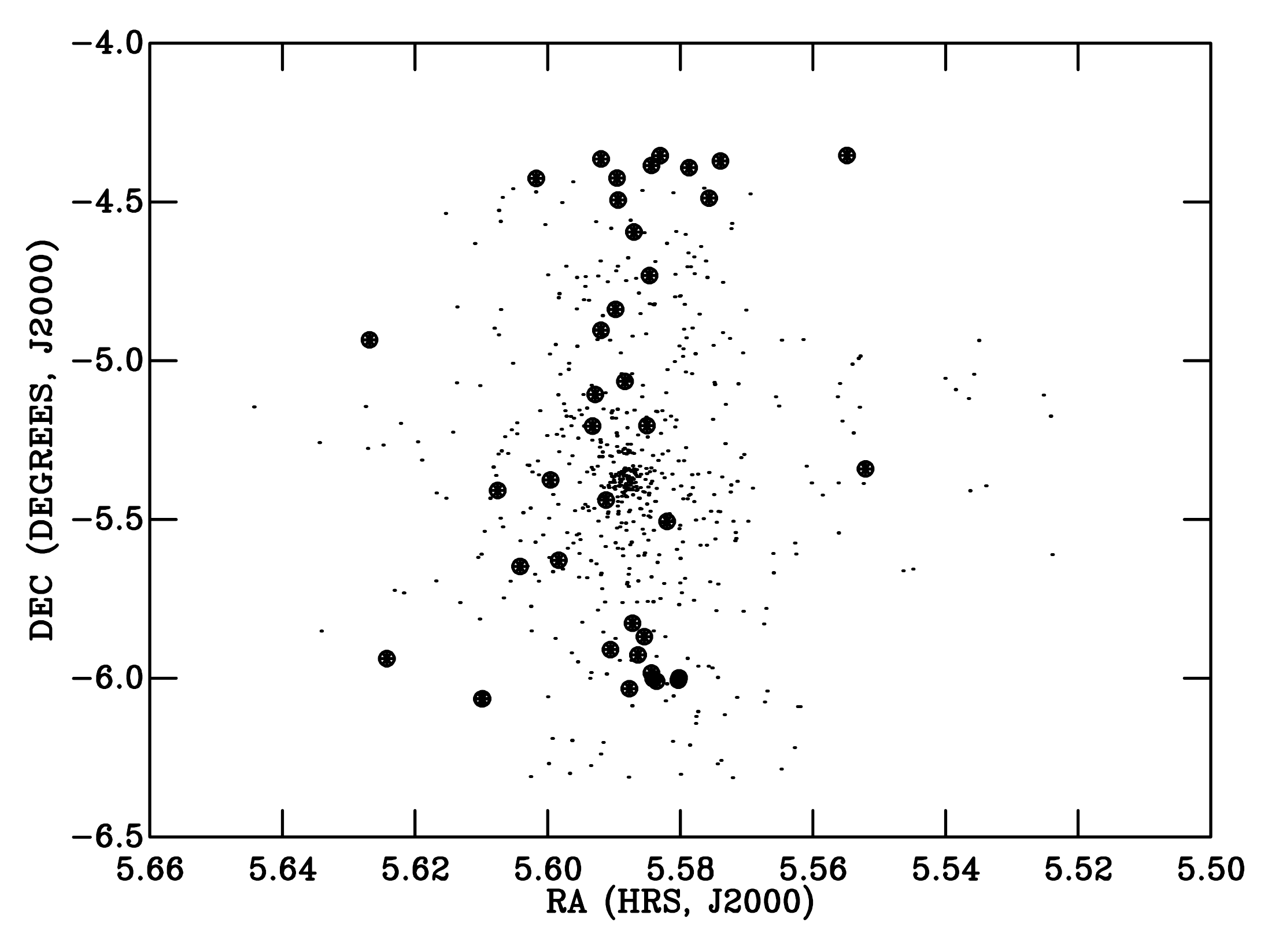}
  \end{center}

  \caption{
The positions of stars in the vicinity of the Orion Nebula Cluster.
The dots in the central region are one in five of the stars from 
Hillendbrand (1997).
The dots around the periphery are X-ray sources from
Ram\'{i}rez et al. (2004). 
The filled circles are the main-sequence sample.
}
  \label{onc_fig}
\end{figure}

The position on the sky of our MS sample is shown in Figure \ref{onc_fig}, along
with the positions of the sample of \cite{1997AJ....113.1733H}, which 
represents stars in the ONC itself, and the flanking fields of
\cite{2004AJ....128..787R}.
Given the distribution of stars, it is clear that the PMS age we should use
is that of the flanking fields.
Although \cite{2004AJ....128..787R} calculate this, they do so on the 
assumption that the ONC is 470pc away, whilst a more modern estimate
is 400pc \citep[][and references therein]{2008MNRAS.386..261M}.
In the $V_0$/$(V-I)_0$ diagram \cite{2004AJ....128..787R} place the flanking 
fields 0.3 mags above NGC2264.
Correcting the distance to 400pc will bring the flanking fields PMS to the same 
magnitude as that of NGC2264, and therefore to an age of 3Myr on the 
scale of \cite{2008MNRAS.386..261M}.

\section{Discussion}
\label{discuss}

\begin{table*}
\caption{Main-sequence and pre-main-sequence ages.}
  \begin{tabular}{@{}lccccclcl@{}}
   Cluster or   & PMS Age & \multicolumn{2}{c}{MS Age (Myr)} & $\Pr(\tau^2)$ & Distance & E(B-V) & Data  & Numbering \\
   Group        & (Myr)   & Best fit   & 68\% confidence     &               & Modulus  &        & table & system reference\\
\\

$\lambda$ Ori & 3   & 6.6 & 5.8--7.5  & 0.52 & 8.05 & 0.12$^1$ & 2 & HD catalogue\\
NGC6530     & 2   & 5.5 & 4.9--6.1  & 0.84 & 10.76 & 0.33$^1$ & 3 & \cite{1957ApJ...125..636W}\\
NGC2264     & 3   & 5.5 & 2.4--6.0  & 0.06 & 9.24 & 0.05 & 4 & \cite{1956ApJS....2..365W}\\
ONC             & 3   & 5.0 & 2.8--5.2  & 0.10 & 7.92 & 0.03 & 5 & \cite{1935POLyo...1...12B}\\
$\sigma$ Ori              & 3   & 0.4 & $<$ 6.6    & 0.26 & 7.98 & 0.06$^2$ & 6 & HD catalogue \\
NGC2362     & 4.5 & 9.1 & 5.4--12   & 0.08 & 10.71 & 0.08 & 7 & \cite{1950ApJ...112..240J}\\
CepOB3b      & 4.5 & 10 & 8.6--10.9 & 0.05 & 8.72 & 0.89$^1$ & 8 & \cite{1959ApJ...130...69B}\\
IC2602         & 25  & 44 & 28--62     & 0.27 & 5.88 & 0.02 & 9 & HD catalogue\\
NGC2547     & 38  & 48 & 27--62      & 0.13 & 8.03 & 0.04 & 10 & \cite{1982AandAS...47..323C}\\
Pleiades       & --  & 115 & 104--117 & 0.81 & 5.35 & 0.02$^1$ & 11 & \cite{1947AnLei..191...1H}\\
\end{tabular}
\\
$^1$Median from individual extinctions.\\
$^2$ From \cite{1994A&A...289..101B}.\\
\label{ages_table}
\end{table*}

\begin{table}
\caption{An sample of Tables 2-11, the fitted dataset for $\lambda$ Ori.
Table \ref{ages_table} gives the number of the electronic table for each dataset, along with the
reference for the star numbering system.  
As shown here, for each cluster we give, along with the uncertainties, the fitted
$V$, $B-V$ and $U-B$.  In the case of groups where extinctions were derived on a star-by-star
basis these are the reddening and extinction free values, and the $E(B-V)$ used is given in the
last column.}
  \begin{tabular}{@{}cccccccccc@{}}
   Star       & \multicolumn{2}{c}{$V$} & \multicolumn{2}{c}{$B-V$} & \multicolumn{2}{c}{$U-B$} & E(B-V)\\
   number     &      mag & $\sigma$     &      mag & $\sigma$     &      mag & $\sigma$      \\
\\
 36822  &   4.036 & 0.010 &  -0.270 & 0.008 &  -1.047 & 0.010 &  0.120\\
 36861  &   3.346 & 0.010 &  -0.271 & 0.008 &  -1.049 & 0.010 &  0.061\\
 36862  &   4.710 & 0.010 &  -0.241 & 0.008 &  -0.953 & 0.010 &  0.281\\
 36894  &   8.662 & 0.010 &  -0.086 & 0.008 &  -0.295 & 0.010 &  0.036\\
 36895  &   6.512 & 0.010 &  -0.188 & 0.008 &  -0.745 & 0.010 &  0.068\\
 37034  &   8.976 & 0.010 &  -0.069 & 0.008 &  -0.216 & 0.010 &  0.109\\
 37035  &   8.249 & 0.010 &  -0.141 & 0.008 &  -0.562 & 0.010 &  0.121\\
 37051  &   8.699 & 0.010 &  -0.074 & 0.008 &  -0.239 & 0.010 &  0.114\\
 37110  &   8.657 & 0.010 &  -0.097 & 0.008 &  -0.346 & 0.010 &  0.097\\
245140  &   8.568 & 0.010 &  -0.107 & 0.008 &  -0.398 & 0.010 &  0.217\\
245168  &   9.077 & 0.010 &  -0.057 & 0.008 &  -0.173 & 0.010 &  0.177\\
245185  &   9.313 & 0.010 &  -0.050 & 0.008 &  -0.153 & 0.010 &  0.190\\
245203  &   6.972 & 0.010 &  -0.188 & 0.008 &  -0.745 & 0.010 &  0.158\\
\end{tabular}
\label{example}
\end{table}

\begin{figure}
  \begin{center}
    \includegraphics[width=3.3in]{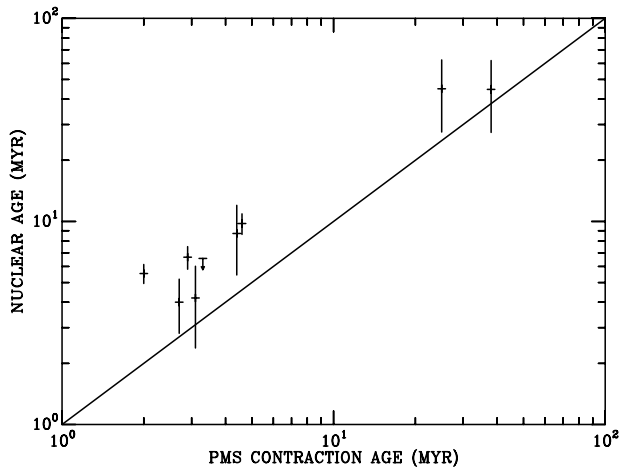}
  \end{center}

  \caption{The main-sequence and pre-main-sequence ages for our sample.
The groups at PMS ages of 3 and 4.5Myr have been separated slightly in age
to aid visibility.}
  \label{ages}
\end{figure}

We collect together our measurements of the ages of the groups and clusters
in Table \ref{ages_table}, along with the other parameters from our
fits.
For completeness we include the distances, though as these are derived from
two-parameter fits we emphasise that those of \cite{2008MNRAS.386..261M}
are to be preferred.
We plot PMS against MS age in Figure \ref{ages}.
Whilst the PMS and MS ages for individual clusters may agree to within the 
uncertainties, the average of the MS ages is significantly older  
than the average of the PMS ages.
If we take only those clusters and associations less than 10Myr old, the
MS ages are, on average, a factor two larger.
The issue is clearly which of these age scales is correct.

\subsection{Are the main-sequence ages incorrect?}
\label{getouts}
 
Explanations as to why the MS ages may be incorrect fall into two groups, 
those associated with the statistical 
techniques and those associated with the models. 
We can rule out problems with the fitting procedure by comparing our ages 
with those obtained by \cite{1993A&AS...98..477M}.
They use similar isochrones to the ones presented here and measure the age 
of the Pleiades as 100 Myr and the environs of the ONC as 4Myr.
Both these ages are compatible with those we measure, suggesting our technique
gives similar ages to ``by eye'' fitting.
Equally importantly, we match the lithium depletion age for the NGC2547 
\citep[34--36Myr][]{2005MNRAS.358...13J}
and are very close to the depletion age for the Pleiades 
\citep[125--130Myr][]{1998ApJ...499L.199S}. 

\begin{figure}
  \begin{center}
    \includegraphics[width=3.3in]{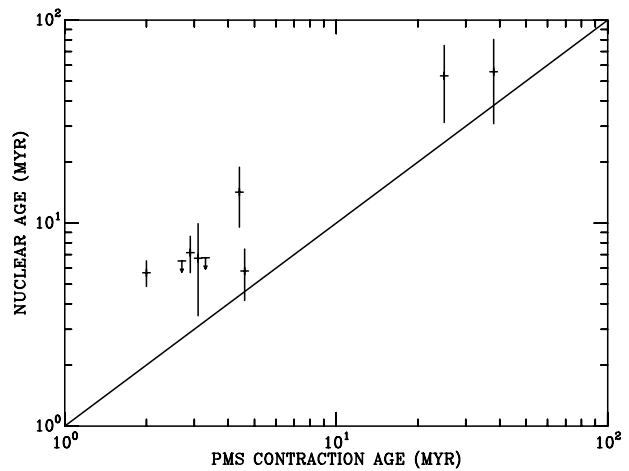}
  \end{center}

  \caption{As figure \ref{ages} but with pre-main-sequence ages calculated
without the brightest star for each dataset.}
  \label{ages_less}
\end{figure}

To check that our uncertainties are at least reasonable we tested how
the result changes if one star is removed from each fit.
As one might expect, the brightest star in the fit provides the tightest 
limits on the age.
We therefore removed the brightest star from each
dataset and replotted Figure \ref{ages}.
As Figure \ref{ages_less} shows, the result remains clear, though as one 
might expect the error bars are larger, and one more dataset (the ONC) 
returns an upper limit for the age.
This strongly suggests that our uncertainty estimates are reasonable, 
and the result is robust.
This experiment also shows what the effect might be of a non-member being 
included in the fit.
Were a non-member very far from the fitted sequence it would have been 
clipped out by the procedure described at the end of Section \ref{the_fit}.
Were it close to the sequence, then it could deviate the fit sufficiently
to have a reasonable $\tau^2$, but then would only change the best fit by
a small amount, similar to the effect of removing a data point.  

As a final check of the uncertainties, in Figures \ref{ngc6530} to \ref{onc}
we plot the data over the best fitting models if the age is fixed at the
PMS age (left) or left as a free parameter (right).
The first two examples (NGC6530 and $\lambda$ Ori) are ones where the PMS age
lies far outside the 68 percent confidence region for the upper-main-sequence
age.
As one expects, we see that the brightest stars lie to the right of the
model when the age is fixed at the PMS age.
Our final example, the ONC, is one where the PMS age lies almost exactly on 
the edge of the 68 percent confidence limit.
Here the improvement in the fit is, as it should be, marginal.
Although such comparisons with our expectations are at best subjective, that
they fit with our expectations adds to our confidence in the result.
When combined with the experiment of missing out the brightest datapoint,
we have a strong case that our uncertainties are correct, and the result
is robust. 

The obvious problems with the models are the absence of rotation, uncertainties
as to the mass-loss rates, and the treatment of convective core overshoot.
Figure 9 of \cite{2000A&A...361..101M} shows that if the stars were rotating, 
and we fitted them with isochrones for stationary stars, the 
resulting ages would be too young by about 10 percent.
This therefore exacerbates the discrepancy between the PMS and MS ages.

All modern models include a degree of core overshoot, which has the effect
of mixing more hydrogen into the core, and hence lengthening the MS
lifetime.
Naively, models with no overshoot will have shorter MS lifetimes than those
used here, by roughly the decrease in available hydrogen (perhaps 
20-40 percent), which is of the right order to bring the MS and PMS ages
back into agreement.
However, our CMD fitting does not measure lifetime on the MS, but how far
from the ZAMS a star at a given luminosity (not mass) has moved.
A close comparison of Figures 4 and 5 of \cite{1981A&A....93..136M} shows that
for the youngest ages they calculate (25Myr) the difference in the position 
of the isochrone corresponds to an age difference of around 5 percent.

Mass-loss rates for early-type stars are uncertain, and so in addition to
using the Geneva models with the standard mass-loss rates 
\citep[set ``c''of ][]{1994A&A...287..803M} we also tried the higher mass-loss
rate, set ``e''.
Comparison of the resulting isochrones for the masses and ages we are 
interested in shows differences in colour which are too small to affect our
results.

Finally, we have tested the effect of using different MS models.
As an alternative to the Geneva models with the \cite{1998A&A...333..231B} 
conversions to colour and magnitude we used the conversions presented
with the isochrones in \cite{2001A&A...366..538L}.
We obtained ages somewhat older than those from the Geneva-Bessell
models, exacerbating the age difference problem.
More importantly, the values of $\Pr(\tau^2)$ are much worse than those 
for the  Geneva-Bessell models, typically around 0.01 or 0.001, showing 
that these models can be ruled out as good descriptions of the data.
To test whether this is the interior models or the atmospheres, we fitted 
the data to the Padova models \citep{2002A&A...391..195G} but with  
the same model atmospheres \citep{1998A&A...333..231B} as we used for the
Geneva-Bessell models.
We find this gives slightly younger ages (a factor 1.5 older than the 
PMS ages in the range 1-10Myr), but very similar values of $\Pr(\tau^2)$ to the
Geneva-Bessell models.
In summary, therefore, our fitting gives strong support for the  
\cite{1998A&A...333..231B} conversions, and there is only a weak effect 
from the interior models, which can explain some, but not all of, the age 
discrepancy.

\subsection{Are the pre-main-sequence ages incorrect?}

The PMS ages are much less robust than the MS ones.
We have adopted the PMS age scale of \cite{2008MNRAS.386..261M}.
However, as \cite{2008MNRAS.386..261M} and \cite{2007MNRAS.375.1220M}
make clear, the primary aim of this scale is an age ordering.
The age scale itself is rather arbitrary, though was chosen to match as
closely as possible the commonly quoted ages for the young groups.
The problem is that there is no single PMS age scale, a point nicely
illustrated in \cite{2009MNRAS.393..538J}.
They show that the $\gamma$ Vel association could have a PMS age between 5 
and 15Myr depending on which PMS models are used, and which part of the 
sequence is considered.
They estimate that the association is about 7Myr old on the 
\cite{2008MNRAS.386..261M} scale, so doubling the ages of these young
associations is consistent with some PMS models.
Our conclusion, therefore, is that the MS age scale is probably the
correct one.

\subsection{Implications of lengthening the PMS timescale}

Before discussing the implications of a longer timescale, we should 
be wary of over interpreting Figure \ref{ages}.
Whilst it clearly shows a discrepancy between mean PMS and MS ages, the
error bars for individual data points are large.
All we can say with any certainty is that there is a difference of
approximately a factor two at PMS ages of 3Myr.
By 30Myr our data are consistent with the age scales matching, though a 
difference of a factor 1.5 is still, in the statistical sense, likely.
We therefore limit ourselves to discussing the implications of a lengthening
of the timescales in the 1-10Myr PMS age range.
Even here, however, we find there are problems it might solve.

There is a long-standing issue that the observed timescale for the
dissipation of proto-stellar discs \citep[3Myr;][]{2001ApJ...553L.153H} 
may be shorter than the time required by the models for planet formation 
\citep[10Myr;][]{1996Icar..124...62P}.
In recent years there has been significant effort to find mechanisms which 
will shorten the planet forming timescales.
Whilst a case can be made that this problem has been solved 
\citep{2008ASPC..398..235M}, there is a view that significant problems remain  
\citep[see, for example, the introductory sections of][]
{2009MNRAS.393...49A, 2008ApJ...688L..99D}.
A fair summary is probably that whilst there are mechanisms which could 
shorten the timescale, such as dust settling \citep{2005Icar..179..415H} 
and planetary migration \citep{2005ApJ...626L..57A}, the uncertainties in 
the physics remain such that it is not clear they do.
Our result offers an interesting alternative solution.
If the clusters used to measure the disc dissipation timescale are 50-100 
percent older than previously thought, there may be no contradiction with the 
\cite{1996Icar..124...62P} timescale.

\cite{2007MNRAS.376..580J} point out that there is a lack of clusters in 
the age range 5--30Myr.
Revising the age scale in the way suggested by the MS fitting would move 
clusters from the youngest ages into this age range.
Furthermore if the age
scales come back into register at around 30Myr, as Figure \ref{ages} 
suggests they might, there would not be a compensating movement out of
the 5-30Myr range, leading to an increased number of clusters at these
ages.

\section{Conclusions}

We have shown that there is a systematic difference between the ages of 
clusters and associations measured from the MS and ages commonly used
which are based on the PMS.
The difference is in the sense that the MS ages are a factor 1.5-2.0
greater than the PMS ages in the age range 2-5Myr (on the PMS scale). 
The most straightforward solution is to adopt the MS age scale, as there
are PMS models which fit with the longer timescale.
Adopting the longer timescale offers a solution to the problem that 
the lifetimes of discs around stars (3-5Myr on the PMS age scale) are
shorter than the time taken to form planets, and to the apparent
absence of clusters in the 5-30Myr age range.

Finally we should be clear that although we favour the age-scale given by MS
fitting, we are not recommending it as a method for deriving ages for 
individual clusters and associations.
As Figure \ref{ages} and Table \ref{ages_table} make clear, the uncertainties 
for individual groups are large.
Nor can we at this point make any clear recommendation as how one should
reflect this result when quoting PMS ages.
Whilst it is clear that the youngest ages need to be increased, how far down
the age scale that should be propagated is unclear.
We therefore continue to commend the Mayne et al/Mayne \& Naylor age ordering, 
though recommend that if these ages are quoted one states clearly 
that they are on the Mayne et al/Mayne \& Naylor scale.
If absolute ages are required for clusters younger than 10Myr for
comparison with other data we recommend multiplying the 
Mayne et al/Mayne \& Naylor values by 1.5 and quoting the age scale as
originating from this paper. 

\section*{Acknowledgments}
I am grateful to three people for provoking significant parts of this work.
First, Herbie (H.D.) Deas for conversations many years ago 
demonstrating that one has choices in mathematics; a view which led 
me to the re-examination of the normalisation presented in Section \ref{norm}.
Second, the (anonymous) referee of \cite{2006MNRAS.373.1251N}, whose efforts 
emphasised to me the inelegance of the method 
presented in that paper for calculating the uncertainties in the parameters, 
and hence led to Section \ref{uncer}. 
Third, the referee of a very early version of this paper forcefully made
the point that it would be improved by the inclusion of data, which led
me to the main result presented here.

This research has made use of the WEBDA database, operated at the 
Institute for Astronomy of the University of Vienna, from where 
much of the electronic form of the data used here orginated.
Finally, the referee of this version of the paper, John Stauffer
prompted several improvements.

\bibliographystyle{mn2e}
\bibliography{bibdesk.bib}

\bsp

\label{lastpage}

\end{document}